# Non-Invasive Diagnosis for Clubroot Using Terahertz Time-Domain Spectroscopy and Physics-Constrained Neural Networks

Pengfei Zhu, *Member*, *IEEE*, Jiaxu Wu, Alyson Deslongchamps, Yubin Zhang, and Xavier Maldague, *Life Senior Member, IEEE*

*Abstract*—Clubroot, a major soilborne disease affecting canola (*Brassica napus* L.) and other cruciferous crops, is characterized by the development of large galls on the roots of susceptible hosts. In this study, we present the first application of terahertz time-domain spectroscopy (THz-TDS) as a non-invasive diagnosis tool in plant pathology. Compared with conventional molecular, spectroscopic, and immunoassay-based methods, THz-TDS offers distinct advantages, including non-contact, non-destructive, and preparation-free measurement, enabling rapid in situ screening of plant and soil samples. Our results demonstrate that THz-TDS can differentiate between healthy and clubroot-infected tissues by detecting both structural and biochemical alterations. Specifically, infected roots exhibit a blue shift in the refractive index in the low-frequency THz range, along with distinct peaks – indicative of disruptions in water transport and altered metabolic activity in both roots and leaves. Interestingly, the characteristic root swelling observed in infected plants reflects internal tissue disorganization rather than an actual increase in water content. Furthermore, a physics-constrained neural network is proposed to extract the main feature in THz-TDS. A comprehensive evaluation, including time-domain signals, amplitude and phase images, refractive index and absorption coefficient maps, and principal component analysis, provides enhanced contrast and spatial resolution compared to raw time-domain or frequency signals. These findings suggest that THz-TDS holds significant potential for early, non-destructive detection of plant diseases and may serve as a valuable tool to limit their spread in agricultural systems.

*Index Terms*—Terahertz time-domain spectroscopy, clubroot, plant disease, canola, deep learning, feature extraction.

## I. INTRODUCTION

CLUBROOT is a significant soilborne disease of canola (*Brassica napus* L.) and other cruciferous crops [1]. It's reported that the clubroot disease causes a 60-90% yield loss. The disease is associated with the formation of large clubs at the roots of susceptible hosts. This disease is caused by the obligate rhizarian protist *Plasmodiophora brassicae* [2].

Severe club development reduces the ability of plants to extract sufficient water and nutrients from the soil, resulting in the stunting of aboveground plant parts [2].

Efficient disease management begins with a reliable and sensitive diagnostic method to prevent the spread of the pathogen and mitigate its economic losses [3]. Gossen et al. [4] compared five molecular techniques for estimating clubroot resting spores in soil, including quantitative polymerase chain reaction (qPCR), competitive positive internal control PCR (CPIC-PCR), propidium monoazide PCR (PMA-PCR), droplet digital PCR (ddPCR), and loop-mediated isothermal DNA amplification (LAMP). Feng et al. [5] combined hyperspectral imaging (HSI) and a convolutional neural network (CNN) to detect the clubroot. The study demonstrates the feasibility of cabbage clubroot detection by phenotyping of aboveground parts using near-infrared HSI combined with CNN. Wakeham and White [6] evaluated the potential of several immunoassays for detecting clubroot dormant in soil, including western blotting, dip-stick, dot-blot, indirect enzyme-linked immunosorbent assay (ELISA), and indirect immunofluorescence techniques. Recently, Feng et al. [10] used magnetic resonance imaging (MRI) technology to detect canola clubroot. They developed classification models and demonstrated that MRI could effectively detect clubroot disease in a non-invasive manner. However, these detection methods offer certain sensitivity but often suffer from limitations, including long processing times, high costs, and difficulty in real-time monitoring.

Therefore, there is an urgent need for a rapid, non-destructive, and highly sensitive detection approach for the early diagnosis of clubroot disease. As a novel technique in this century, terahertz time-domain spectroscopy (THz-TDS) has gained interest in analytical chemistry [7], material characterization [8], non-destructive testing [9], [10], etc. Terahertz radiation is a general term for electromagnetic radiation in the broadband, which is located between microwave and infrared, with the frequency ranging from 0.1

This work was supported in part by the Natural Sciences and Engineering Research Council of Canada (NSERC) through the CREATE-oN DuTy! Program under Grant 496439-2017, in part by the Canada Research Chair in Multi-polar Infrared Vision (MIVIM). The Fonds de recherche du Québec – Nature et technologies (Grant number: 350513) provides doctoral research scholarship to J.W. (Corresponding author: Pengfei Zhu. Co-first authors: Pengfei Zhu and Jiaxu Wu.)

Pengfei Zhu, Yubin Zhang, and Xavier Maldague are with the Department of Electrical and Computer Engineering, Computer Vision and Systems Laboratory (CVSL), Laval University, Québec G1V 0A6, Québec city, Canada (e-mail: pengfei.zhu.1@ulaval.ca; zzhyubin@163.com; xavier.maldague@gel.ulaval.ca).

Jiaxu Wu is with the Department of Plant Sciences, Laval University, G1V 0A6, Québec city, Canada (e-mail: jiaxu.wu1@ulaval.ca).

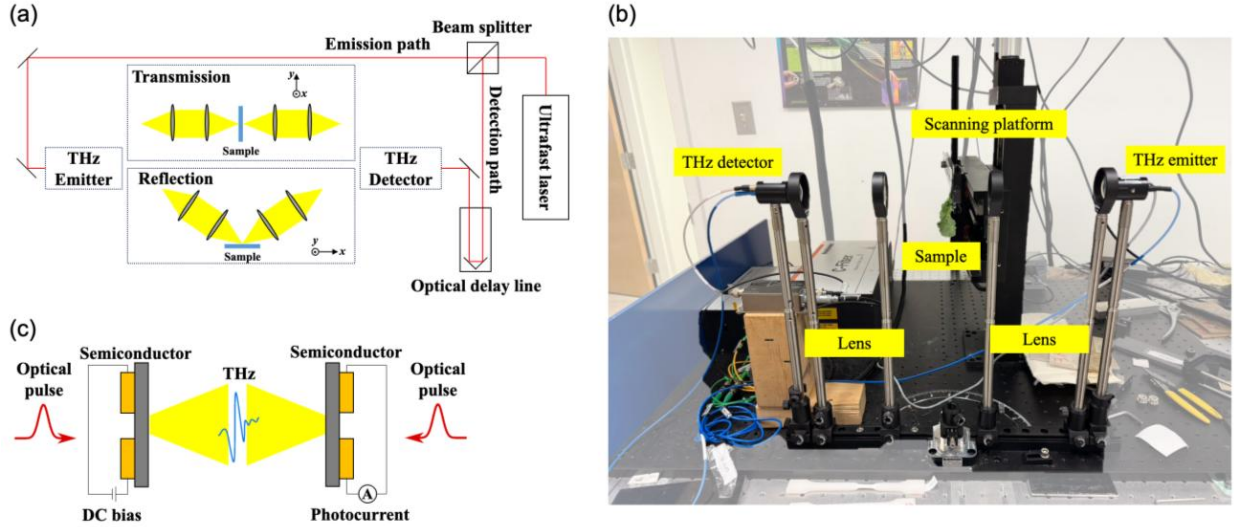

**Fig. 2.** Experimental setup: (a) Schematic image of THz-TDS system; (b) Photograph of THz-TDS system; (c) Work principle of photoconductive antenna.

THz to 10 THz. Mid-infrared, near-infrared, and Raman spectroscopy treat high-frequency vibrations assigned to groups in molecules. In contrast, THz spectroscopy can determine the low-frequency vibrational modes of molecules, which are related to intermolecular interactions in chemical compounds [11], such as hydrogen bonds and van der Waals interactions between molecules. Most common amino acids (beta-form crystal of L-glutamic acid) are composed of hydrogen bonds between molecules, which will exhibit THz peaks resonated with hydrogen bonds in the crystals [12]. According to this mechanism, molecular network information can be obtained, such as high-order protein structures and DNA double strands [13]. Biological molecules in water tend to form clusters, which are very important in the life sciences and biotechnology fields because it largely determines the effectiveness of medications and the high-order protein structures [14]. However, there is still not fully understood because cluster size is in the nanometer order. It has been known for some time that the resonant frequencies of these molecular networks are in the THz range, which makes THz an ideal tool for revealing the molecular networks such as organic acids [15], amino acids [16], sugars [17], polypeptides [18], DNA [19], proteins [20]. However, in the field of plant science, most research has focused on characterizing moisture and drought stress [21], [22] [23], [24], [25]. It is urgent to further extend the THz-TDS technique to quantitatively characterize the resistance to plant diseases.

In this work, we introduce a novel approach for detecting clubroot using THz-TDS. To the best of our knowledge, this is the first application of THz technology for clubroot detection. A comprehensive spectral analysis was conducted to distinguish between infected and healthy plants. Initially, *Arabidopsis thaliana* (hereafter: *Arabidopsis*), a widely used model organism in plant research, which is also infected by *P. brassicae*, was tested. Subsequently, THz-TDS was used to analyze both healthy and infected canola. Notably, the analysis and discussion were structured around the root and leaf components. Unlike existing work in biological fields, the physical properties of plants, including refractive index and absorption coefficient, were extracted to analyze variations in structural and water content. Finally, to effectively and accurately extract the feature of THz-TDS, a physics-constrained neural network is proposed. The important optical parameters (refractive index and absorption coefficient) are embedded into the deep autoencoder. The results are compared to those of conventional feature extraction algorithms to validate the feasibility and robustness.

## II. EXPERIMENTAL SETUP AND SAMPLES

### A. Plant Materials and Growing Conditions

The *Arabidopsis* Col-0 ecotype and *B. napus* Westar variety were selected in this study. *Arabidopsis* seedlings were grown in a climate-controlled growth chamber at 21 °C under a 16-h photoperiod with 60% relative humidity. Canola seedlings were grown in a climate-controlled greenhouse at 24 °C under an 18-hour photoperiod with the same humidity conditions.

### B. P. brassicae Infections

*P. brassicae* 3A isolate was used in this study [26]. One milliliter of $1.0 \times 10^8$ spores/mL resting spore suspension was inoculated onto 14-day-old *Arabidopsis* Col-0 seedlings and 7-day-old canola seedlings through the soil around the roots. The clubroot phenotype evaluation was followed by 21 days post-inoculation (dpi) for *Arabidopsis* and 28 dpi for canola.

### C. Terahertz Time-Domain Spectroscopy

A photoconductive antenna (PCA) comprising two metal electrodes deposited onto a semiconductor substrate was used to generate and detect THz pulses, as shown in Fig. 2(c). To generate THz, the antenna gap was voltage-biased and illuminated with a femtosecond laser pulse (Fig. 2(a)). The rapid rise and fall of the transient photocurrent switching could generate broadband THz pulses with bandwidths up to 15 THz. To improve the signal-to-noise ratio (SNR), the emitter antenna was biased with an alternating voltage and lock-in amplification was used. For THz detection, the excited photocarriers are accelerated by the incoming THz radiation, leading to a time-varying photocurrent $J(\omega)$ between the antenna electrodes. The ultra-fast laser pulse is split into a pump beam and a reference beam. The pump beam is time-shifted using an optical time-delay line and then adapted to excite a THz pulse using a THz emitter. The THz wave penetrates through the samples to a coupled detector. The reference beam is implemented here on the detector as the sampling signal. The THz signal after sampling is transferred to a lock-in amplifier. Amplification of the signal is performed for data acquisition, as shown in Fig. 2(b).

The THz-TDS system (Menlo Systems) is configured in transmission modes to obtain the time domain spectrums of samples, as shown in Fig. 2(a) and Fig. 2(c). Two freely tunable lasers operate at wavelengths of 1560 nm and 780 nm, with total average output powers exceeding 500 mW and 250 mW, respectively. The pulse durations are less than 90 fs and 100 fs for the two lasers. The fiber-coupled terahertz spectrometer features a bandwidth exceeding 6 THz, a dynamic range greater than 100 dB (up to 110 dB), an average THz power up to 300 μW, a scan range of up to 1700 ps, and a spectral resolution of less than 0.6 GHz. The scanning step is set at 0.5 mm. During the measurement period, the facilities were monitored by an air conditioning system that constantly adjusted the temperature to within 21 °C ± 0.5 °C.

## III. PHYSICS-CONSTRAINED NEURAL NETWORK

In this work, a novel physics-constrained neural network (PCNN) is proposed and applied to THz-TDS. The PCNN is the combination of a deep autoencoder and optical properties measurement in transmission mode (including refractive index and absorption coefficient).

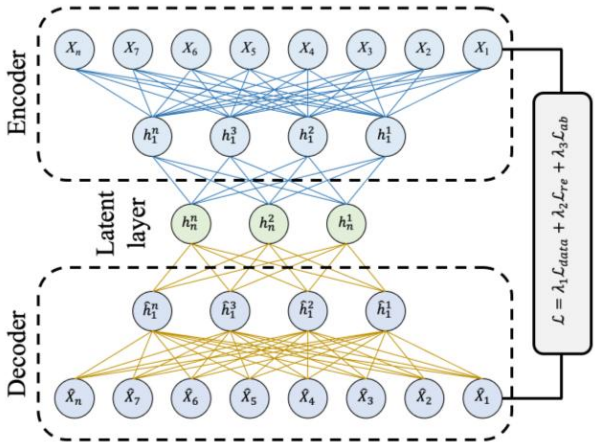

**Fig. 3.** The architecture of the proposed PCNN.

### A. Refractive Index and Absorption Coefficient

Generally, optical parameter measurement is performed in transmission to avoid the impact of alignment accuracy. The refractive index $n(\omega)$ in transmission mode could be calculated based on the phase difference between the sample signal and the reference signal:

$$n(\omega) = 1 + \frac{c}{2\pi\omega d}(\varphi_{sam}(\omega) - \varphi_{ref}(\omega)) \tag{1}$$

where $\varphi_s(\omega)$ and $\varphi_{ref}(\omega)$ are the phase angles of the sample signal and reference signal, $c$ is the speed of the light, $\omega$ is the frequency, and $d$ is the sample thickness. The absorption coefficient can be calculated as:

$$\alpha(\omega) = -\frac{2}{d}\ln\left[r(\omega)\frac{(n(\omega)+1)^2}{4n(\omega)}\right] \tag{2}$$

where $r(\omega) = E_{sam}(\omega) / E_{ref}(\omega)$ is the ratio between sample signal $E_{sam}(\omega)$ and reference signal $E_{ref}(\omega)$.

### B. Physics-Constrained Neural Network

An autoencoder is an unsupervised neural network architecture that learns compact latent representations of input data in an efficient and data-driven manner. Its primary objective is to extract low-dimensional yet informative features that preserve the essential characteristics of the original signal. As illustrated in Fig. 3, the autoencoder is composed of two fundamental modules: an encoder and a decoder.

The encoder transforms the input data $X \in \mathbb{R}^n$ into a lower-dimensional latent vector from $h \in \mathbb{R}^m$ $(m < n)$, retaining the most representative information while discarding redundancy. The encoding operation for the first layer can be formulated as:

$$h_1 = f_1(\omega_1 X + b_1) \tag{3}$$

where $\omega_1$ denotes the weight matrix connecting the input layer to latent layers, $b_1$ represents the bias term, and $f(\cdot)$ is a nonlinear activation function. For deeper layers, the encoding process generalizes to:

$$h_i = f_i(\omega_i h_{i-1} + b_i) \tag{4}$$

where $i$ denotes the layer number.

The decoder performs the inverse mapping, reconstructing the input from the latent representation $h$ back into the original feature space:

$$\hat{h}_{L-i} = g_{L-i}(\omega'_{L-i}\hat{h}_{L-i+1} + b'_{L-i}) \tag{5}$$

where $L$ is the total number of encoder layers, and $\omega'$ and $b'$, and $g(\cdot)$ denote the decoder's weights, biases, and activation functions, respectively. The final reconstruction layer yields:

$$\hat{X} = g_1(\omega'_1\hat{h}_1 + b'_1) \tag{6}$$

where $\hat{X}$ represents the network's reconstructed output. Ideally, $\hat{X}$ should closely approximate the original input $X$. To achieve this, the encoder and decoder parameters are jointly optimized by minimizing a reconstruction loss, commonly defined as the mean-square error (MSE):

$$\mathcal{L}_{data} = \frac{1}{N}\sum_{i=1}^{N}||X_i - \hat{X}_i||_2^2 \tag{7}$$

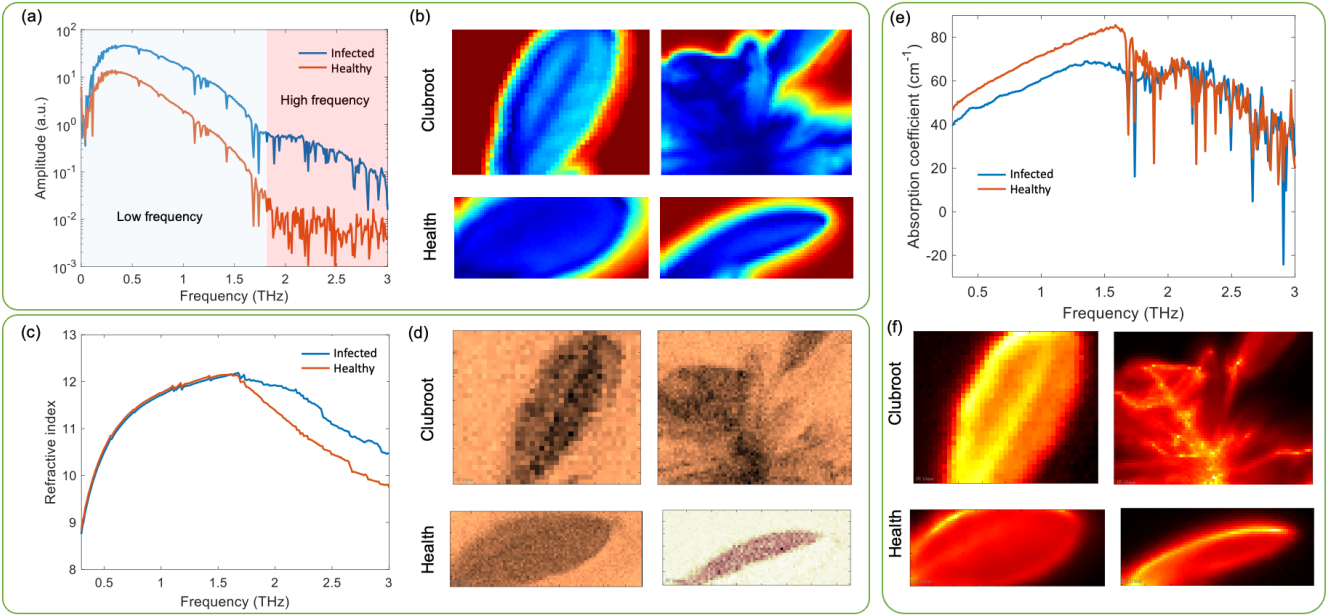

**Fig. 4.** Experimental results from *Arabidopsis*: (A) THz amplitude spectrum, (B) THz amplitude images, (C) Refractive index curves, (D) Refractive index maps, (E) Absorption coefficient curves, (F) Absorption coefficient maps.

Despite the development of several variants—including denoising [27], sparse [28], contractive [29], and variational autoencoders [30]—most still encounter limitations in representing complex data distributions or adapting to diverse application scenarios when trained solely using $\mathcal{L}_{data}$.

For THz-TDS feature extraction, if only the data-driven loss defined in Eq. (7) is employed, the neural network primarily captures correlations inherent to the dataset rather than learning the physical characteristics intrinsic to the THz spectral range. Consequently, its feature representations may resemble those obtained by conventional principal component analysis (PCA). In some cases, interference patterns can further obscure the intrinsic spectral signatures, leading to limited interpretability. To address this issue, a physics-guided constraint is incorporated into the deep autoencoder framework to enforce physical consistency in the learned features.

The overall architecture of the terahertz physics-guided neural network (THz-PGNN) is illustrated in Fig. 3. In addition to the data reconstruction loss $\mathcal{L}_{data}$, a physics-based loss is introduced to constrain the equivalence of the refractive index and absorption coefficient between the input and reconstructed spectra:

$$\mathcal{L}_{phys} = \mathcal{L}_{re} + \mathcal{L}_{ab} \tag{8}$$

$$\mathcal{L}_{re} = \frac{1}{N}\sum_{i=1}^{N}(\frac{c}{2\pi\omega d}(\varphi_{sam}^0(\omega) - \varphi_{ref}^0(\omega)) - \frac{c}{2\pi\omega d}(\varphi_{sam}^1(\omega) - \varphi_{ref}^1(\omega)))^2 \tag{9}$$

$$\mathcal{L}_{ab} = \frac{1}{N}\sum_{i=1}^{N}(-\frac{2}{d}\ln\left[r^0(\omega)\frac{(n^0(\omega)+1)^2}{4n^0(\omega)}\right] + \frac{2}{d}\ln\left[r^1(\omega)\frac{(n^1(\omega)+1)^2}{4n^1(\omega)}\right])^2 \tag{10}$$

where the superscripts 0 and 1 correspond to the input and reconstructed signals, respectively. The total loss function thus combines the data fidelity and physics-consistency terms as

$$\mathcal{L} = \lambda_1\mathcal{L}_{data} + \lambda_2\mathcal{L}_{re} + \lambda_3\mathcal{L}_{ab} \tag{11}$$

where $\lambda_1$, $\lambda_2$, and $\lambda_3$ are the weighting coefficients that balance the contributions of the individual terms. It is worth noting that the reference signal, obtained from an aluminum plate, is used solely for physical calibration and is not directly involved in either the input or the reconstruction of the autoencoder.

The architecture consists of an encoder–decoder structure with four convolutional layers in the encoder, progressively reducing the temporal dimension while increasing the feature channels ($1 \rightarrow 16 \rightarrow 32 \rightarrow 64 \rightarrow 128$), followed by an adaptive average pooling layer and a fully connected layer that compresses the representation into a 32-dimensional latent space. The decoder mirrors this structure with one fully connected layer and four transposed convolutional layers to reconstruct the signal back to its original length. Linear interpolation to ensure temporal alignment.

TABLE I
TRAINING PARAMETERS.

| Item | Description |
|---|---|
| Batch size | 32 |
| Epochs | 50 |
| Initial learning rate | $1 \times 10^{-3}$ |
| Input size | [32, 1, 3072] |
| Loss function | Eq. (11) |
| Physics weight $\lambda(t)$ | Linearly increases after 10th epoch, up to 1.0 |
| Scaling factor ($S$) | 1000.0 |
| Optimizer | Adam |
| Training noise level | 25 |
| Gradient clipping | Max norm = 5.0 |

The network is trained using a combined loss function that includes both the mean squared error (MSE) for signal reconstruction and a physics-based constraint term derived from the frequency-domain reflection ratio and phase difference between the reconstructed and reference signals. This physics loss enforces smoothness in the frequency-dependent refractive index $n(\omega)$ and absorption coefficient $\alpha(\omega)$, thus embedding physical consistency into the learning process. The total loss is expressed as $L_{total} = L_{recon} + \lambda(t) \cdot L_{phys}/S$, where $S = 1000$ and $\lambda(t)$ increases linearly after the 10th epoch to balance the physical constraint. The model is trained using the Adam optimizer with a learning rate of $1 \times 10^{-3}$, batch size of 32, and a total of 50 epochs. Gradient clipping (max norm 5.0) is applied to stabilize the training process. This hybrid learning framework effectively combines data-driven representation learning with physics-based regularization, achieving improved signal reconstruction fidelity and physically interpretable latent representations. The details of training parameters are shown in Table 1.

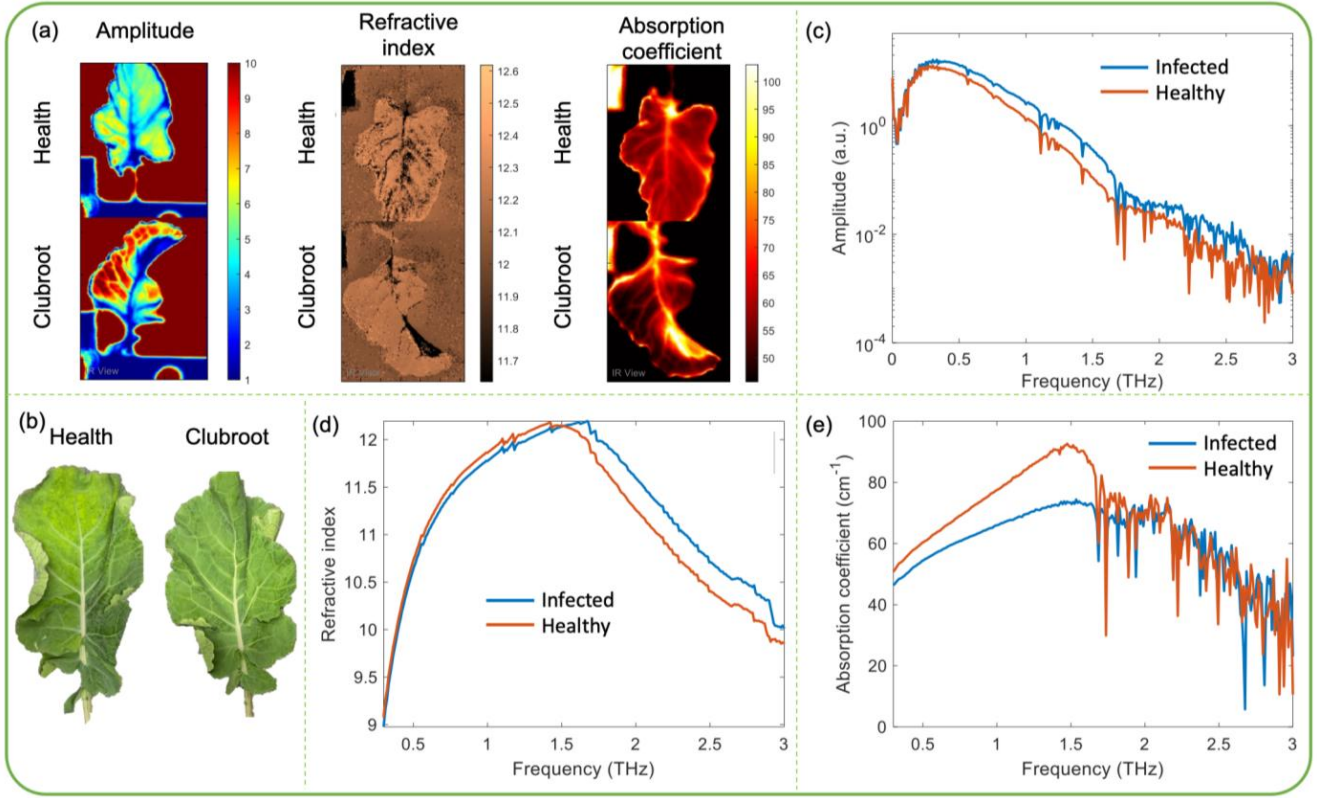

**Fig. 5.** Experimental results from *Brassica napus*: (A) THz amplitude images, refractive index maps, and absorption coefficient maps, (B) Photograph of tested specimens (health and clubroot), (C) THz amplitude spectrum, (D) Refractive index curves, (E) Absorption coefficient curves.

## IV. RESULTS AND DISCUSSION

### A. Spectral Analysis – Arabidopsis

Given that clubroot primarily targets plant tissues, our analytical approach was divided into two complementary components: root-based detection and leaf-based detection. Root analysis offers direct insights into spectroscopic alterations associated with pathogen-induced damage, while leaf analysis evaluates the feasibility of employing THz-TDS as a truly non-invasive and non-destructive diagnostic tool. Prior to applying THz-TDS to *B. napus*, we used the model plant *Arabidopsis* to characterize spectral changes associated with clubroot infection and to establish foundational parameters for subsequent detection in crop species.

We first conducted THz-TDS measurements on leaf tissues. To ensure experimental consistency and comparability, all measurements were taken from anatomically similar positions in more than 20 plant samples. As shown in Fig. 4(a), a marked difference in spectral amplitude was observed between clubroot-infected and healthy plants. Infected plants exhibited notably lower amplitude, particularly in the 0.5-1.5 THz range, indicating higher absorption, whereas healthy plants showed higher amplitude values, likely due to greater and more uniform internal water content. Further examination of the THz images (Fig. 4(b)) revealed that water was more evenly distributed across the leaf surface in healthy *Arabidopsis plants*, whereas in infected samples, the vein structures became clearly delineated. This suggests that clubroot infection alters the distribution of water and physiological properties within the leaf. These changes may be attributed to hormonal dysregulation and vascular remodeling triggered by the pathogen, resulting in increased water retention in the veins, reduced water content in interveinal regions, and structural modifications such as thickened cell walls and elevated vascular density.

Additionally, the spectral features associated with clubroot infection were analyzed (Fig. 4(a)). In the low-frequency range, spectral signatures primarily reflected water-related absorption ("water peaks"). However, at higher frequencies, the spectra became more complex. Healthy plants exhibited lower amplitude and signal-to-noise ratios, whereas infected samples displayed enhanced amplitude with distinct feature peaks, further supporting the presence of clubroot-induced structural and biochemical alterations. These findings demonstrate that clubroot infection has a significant impact on leaf vein architecture and water distribution in *Arabidopsis*. Root tissues were not examined in this species due to their sub-millimeter lateral dimensions, which fall below the spatial resolution (diffraction limit) of the THz-TDS system, rendering such measurement unreliable [31].

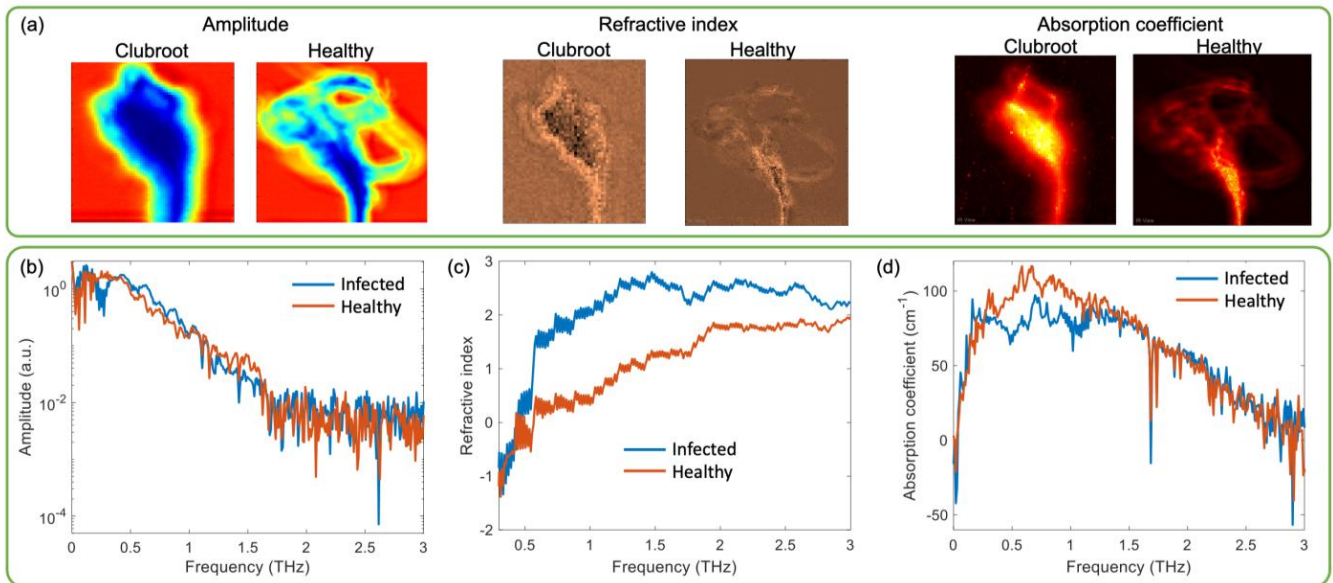

**Fig. 6.** Experimental results from Brassica napus at roots: (A) THz amplitude images, refractive index maps, and absorption coefficient maps, (B) THz amplitude spectrum, (C) Refractive index curves, (D) Absorption coefficient curves.

Previous studies in the THz and plant biology fields have primarily focused on analyzing time- and frequency-domain signals, often overlooking changes in underlying physical parameters. However, key optical properties such as the refractive index and absorption coefficient are particularly informative in the terahertz range, where electromagnetic (EM) waves interact strongly with molecular rotations, vibrations, and hydrogen bonding dynamics.

As shown in Fig. 3(c), clubroot-infected plants exhibit a slight blueshift in the refractive index curve. Although the primary site of infection is the root, the pathogen induces systemic physiological responses that manifest in the leaves, affecting their water content, structural integrity, and chemical composition. Notably, the refractive index of healthy leaves is higher than that of infected leaves in the low-frequency region (< 1.6 THz) but becomes lower in the high-frequency region (> 1.6 THz). This frequency-dependent reversal reflects distinct EM interaction mechanisms: at low frequencies, the EM response is dominated by water content – specifically, hydrogen bonding – while at higher frequencies, it is influenced by accumulation of polar biomolecules such as proteins, phenolics, and flavonoids. These findings align with the complex absorption peaks observed in Fig. 3(a). Refractive index maps (Fig. 3(d)) further illustrate that infected specimens exhibit elevated refractive indices in the high-specrency region. Similarly, the absorption coefficient curves (Fig. 3(e)) show minimal variation at low frequencies but reveal distinct absorption peaks at higher frequencies in infected samples. The corresponding absorption coefficient maps (Fig. 3(f)) reveal pronounced vein structures, highlighting altered distributions of water and metabolites. These observations strongly suggest that clubroot infection induces not only localized root damage but also systemic alterations in leaf biochemistry. Given the complexity of these EM responses in leaf tissues, it is reasonable to infer that THz spectral signatures of infected roots – where the pathogen is most concentrated – are likely to be even more pronounced and informative.

### B. Spectral Analysis – B. napus

We next applied THz-TDS to investigate leaf tissues of *B. napus*, as shown in Fig. 5. Similar to the findings in *Arabidopsis*, clubroot infection altered water distribution and leaf vein structure (Fig. 5a). THz amplitude spectra of infected and healthy plants displayed comparable trends (Fig. 5(c)), though the amplitude difference in *B. napus* was less pronounced than that observed in *Arabidopsis*. Notably, these physiological differences were not visually discernible under standard imaging conditions (Fig. 5(b)), highlighting the utility of THz imaging for detecting subtle disease-related changes. Together, the results from Fig. 4 and Fig. 5 demonstrate that THz-TDS can reveal minor amplitude variations and structural differences in leaves associated with clubroot infection, enabling early-stage detection and assessment of plant health status. As shown in Fig. 5(d), infected *B. napus* leaves exhibited a clear blueshift in the refractive index curve, which was more prominent than that observed in *Arabidopsis* (Fig. 4(c)). Consistent with previous results, the refractive index of infected samples was lower than that of healthy controls in the low-frequency region but higher at high frequencies. This shift suggests that clubroot not only disrupts water transport – as reflected by a reduced dielectric response at low frequencies – but also induces metabolic changes that enhance the dielectric properties at higher frequencies, likely due to an increased accumulation of polar biomolecules. These findings are further supported by the absorption spectra shown in Fig. 5(e).

Following the leaf-based analysis, we conducted THz-TDS measurements on root tissues of *Brassica napus*. As shown in Fig. 6(a), clubroot infection induced significant morphological alterations, including swelling and gall formation. Water and nutrient accumulation appeared concentrated near the central of the infected roots. These spatial changes were clearly reflected in the refractive index and absorption coefficient maps in Fig. 6(a), where infected specimens exhibited disrupted transport patterns, while healthy roots displayed continuous and uniform distribution of water and nutrients. Amplitude spectra (Fig. 6(b)) revealed substantial differences between infected and healthy roots, with clear spectral divergence at both low and high frequency ranges – unlike the leaf spectra shown in Fig. 3 and Fig. 4, where differences were more localized. Notably, a strong absorption band appeared around 0.4 THz in infected roots. Refractive index analysis (Fig. 6(c)) showed that infected roots consistently exhibited lower refractive indices than healthy counterparts across the entire THz band, in contrast to the frequency-dependent trends observed in leaves (Fig. 4(c) and Fig. 5(d)). This decline in refractive index is indicative of reduced water content, as water strongly influences dielectric properties in the THz range. Typically, the refractive index of dry plant tissue lies between 1.2 and 1.6, whereas water-rich tissues exhibit values between 2.1 and 2.4. Thus, the lower refractive index observed in infected roots supports the conclusion that *Plasmodiophora brassicae* infection impairs vascular function and induces tissue dehydration.

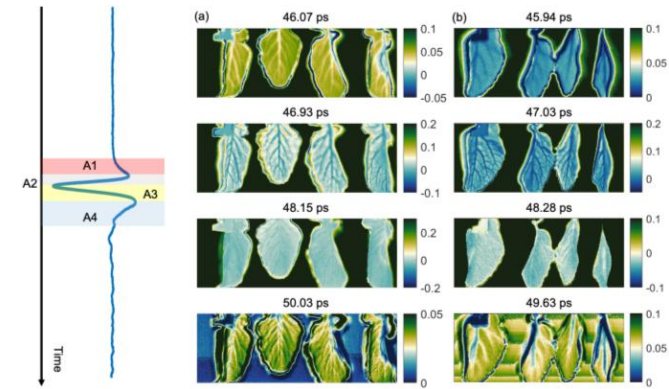

**Fig. 7.** Feature extraction based on time-domain signals: (a) healthy plant; (b) with clubroot.

Clubroot pathogenesis involves rapid cell division, abnormal cell enlargement, vacuolation, and eventual cellular disintegration, all of which contribute to uneven water distribution, tissue loosening, and softening [32]. Therefore, the observed root "swelling" represents structural deformation rather than an increase in water content. Absorption coefficient spectra (Fig. 6(d)) further underscore the contrast between leaf and root responses. While leaf spectra (Fig. 4(e) and Fig. 5(e)) showed smooth, monotonic increases dominated by water absorption, the root spectra appeared more irregular, particularly in infected samples. This difference arises from the inherently more heterogeneous anatomy of roots, which comprises xylem, phloem, cortex, and pericycle cells. Upon infection, excessive cell proliferation, necrosis, vacuolation, and enlarged intercellular spaces lead to pronounced structural disorder. Moreover, weak but distinct resonance absorption features were observed in the low-frequency region, likely originating from organic compounds and secondary metabolites accumulated in response to infection. These findings suggest that THz spectral features in roots offer high sensitivity to structural and compositional changes during disease progression.

### C. Feature Extraction

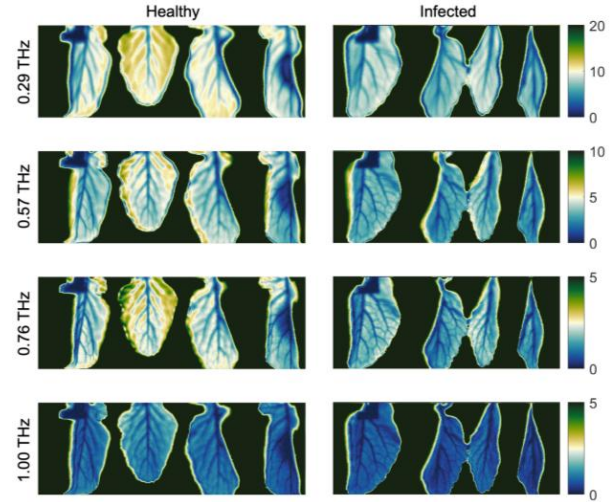

**Fig. 8.** Feature extraction based on amplitude images.

In this section, various feature extraction modalities were employed, including time-domain signals, amplitude and phase images, refractive index and absorption coefficient maps, principal component analysis (PCA), conventional autoencoder, and the proposed physics-constrained neural network (PCNN). The time-domain signals were extracted according to the analysis in our previous work [33]. The time-domain signals were divided into four regions, "A1", "A2", "A3", and "A4", according to the peak and trough, as shown in Fig. 7. In the "A1" region, healthy plants are obviously different from the plants with clubroot. In the "A2" region, the leaf veins are clear in both healthy and diseased plants. The healthy plants exhibit higher THz intensity. However, in the "A4" region, the results are opposite. In the "A3" region, the images become blurred. Therefore, time-domain signals are difficult to identify healthy and infected plants.

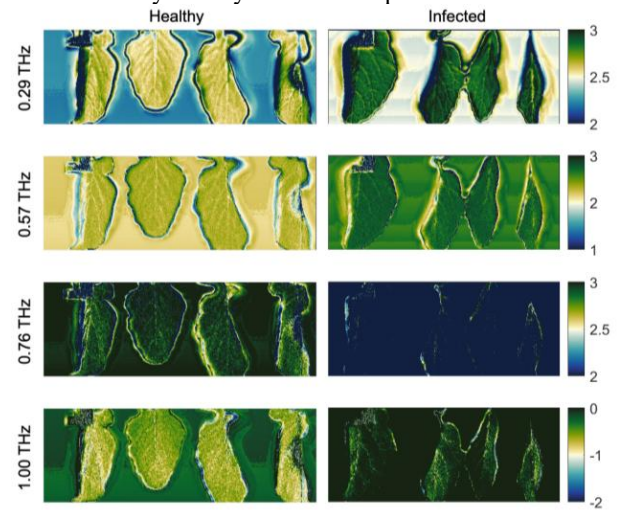

**Fig. 9.** Feature extraction based on phase images.

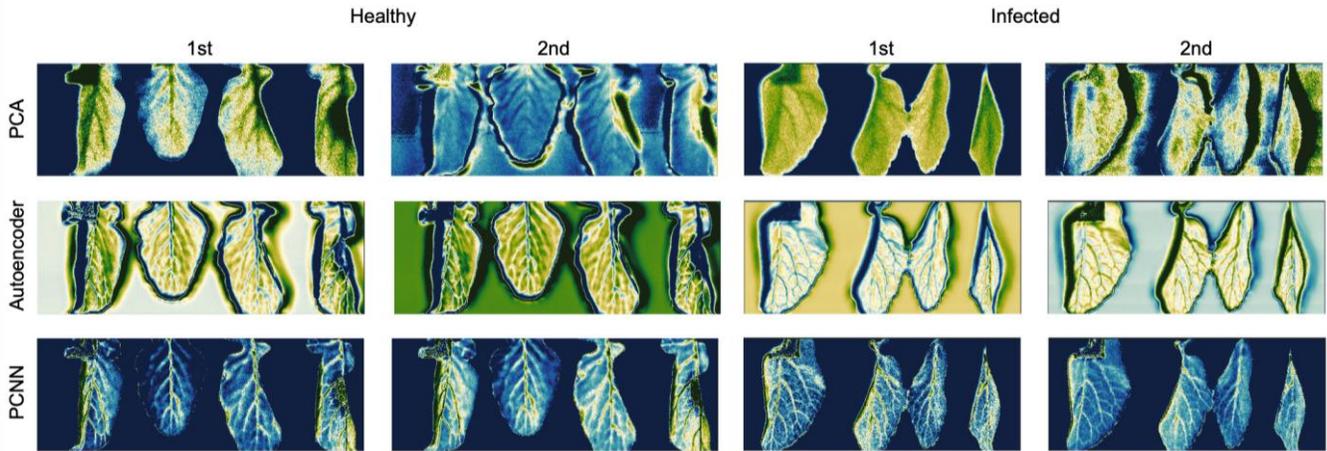

**Fig. 12.** Feature extraction based on PCA, autoencoder, and PCNN methods.

The amplitude images are shown in Fig. 8. In the selected frequencies, the amplitude of healthy plants is higher than that of plants with clubroot. The results are the same as the results in the previous sections. However, in amplitude images, some healthy plants exhibit similar amplitude compared with plants with clubroot (see the fourth leaf). In addition, with the increase in frequency, the distinction between healthy and diseased plants becomes minor.

The phase images are shown in Fig. 9. Different from amplitude images, phase images present a larger distinction. At each frequency, the phase values of healthy plants are far from those of diseased plants. With the increase in frequency, the distinction between healthy and diseased plants becomes major. However, the phase images provide limited texture features on the leaves.

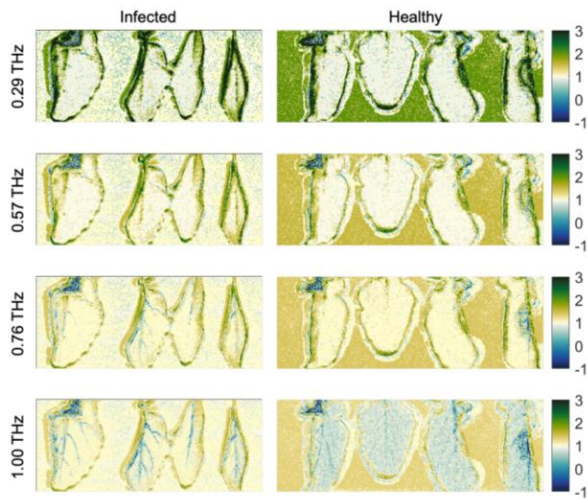

**Fig. 10.** Feature extraction based on refractive index maps.

The refractive index maps are shown in Fig. 10. Except for 0.76 THz and 1.00 THz, it is hard to find the difference between healthy and diseased plants. The absorption coefficient maps are shown in Fig. 11. Comparing with the amplitude images, the absorption coefficient maps show similar results. The distribution of leaves is clear, especially for the leaf veins. The absorption coefficient of healthy coefficient is higher than that of plants with clubroot.

Then, as a commonly used unsupervised feature extraction method, the principal component analysis (PCA) is employed in this work, as shown in Fig. 12. Based on 1st principal component (PC), it is possible to observe the distinction between the healthy and diseased plant for the second leaf. However, the image contrast of other leaves is weak. In particular, for the 2nd PC, the images become blurred. The PCA method cannot differentiate healthy and diseased plants.

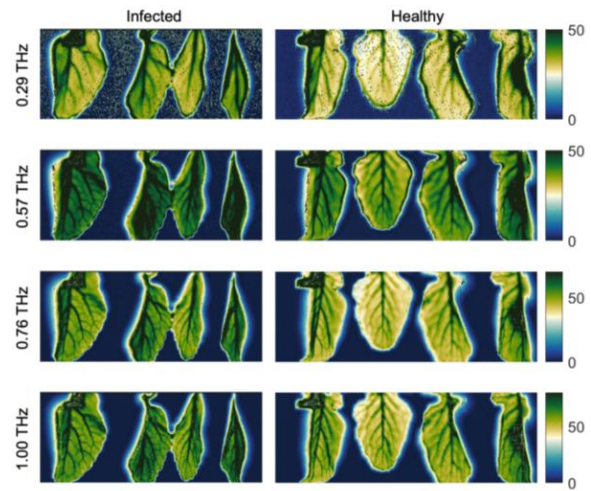

**Fig. 11.** Feature extraction based on absorption coefficient maps.

To compare with the proposed PCNN, we also trained a conventional autoencoder without introducing physical constraints. The results are shown in Fig. 12. Notably, both the autoencoder and PCNN are trained on the non-infected four plants. Since this is a self-supervised learning technique, we do not need to collect too many datasets for training. In the autoencoder's results, the texture of leaves is more obvious than that of the PCA methods. However, there is still an unclear distribution on the leaf, which interferes with us to differentiate healthy and diseased plants.

Finally, the proposed PCNN is employed on these plants, as shown in Fig. 12. Compared with results processed by the previous techniques, the PCNN achieves the best image contrast. It is obvious to observe the leaf veins on both healthy and diseased plants. In addition, we do not need to artificially select suitable images for comparing as the first and second latent are sufficient to explain the main features of the tested samples. Although the proposed PCNN can extract clear features of both healthy and diseased plants, it remains challenging to identify them.

The reason is that we didn't combine them and map them onto the same color map. Then, the results of 1st latent feature on the same colormap are shown in Fig. 13. It is clear to find that healthy plants present a dark color while diseased plants present a bright color.

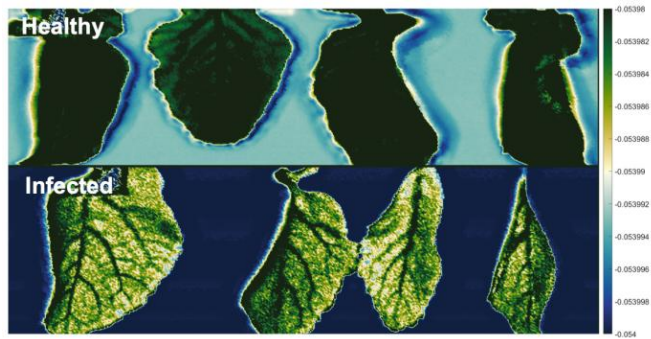

**Fig. 13.** The 1st latent feature extracted by the proposed PCNN, where both healthy and diseased plants are mapped onto the same colormap.

We performed the same processing for the second latent feature extracted by the proposed PCNN. The results are shown in Fig. 14. Obviously, it is quite easy to identify healthy and diseased plants.

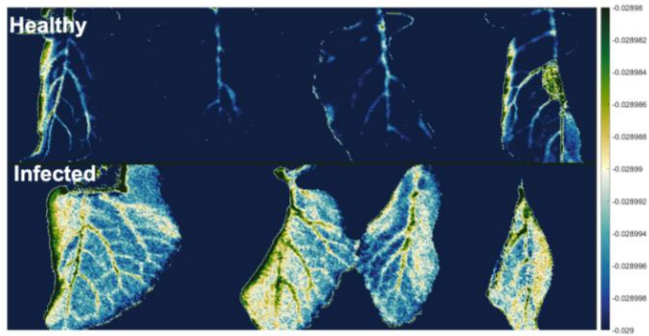

**Fig. 14.** The 2nd latent feature extracted by the proposed PCNN, where both healthy and diseased plants are mapped onto the same colormap.

## V. CONCLUSION

In this paper, we present the first comprehensive application of terahertz time-domain spectroscopy (THz-TDS) as a non-invasive diagnosis tool in plant pathology. Unlike previous work that primarily focused on visualizing water distribution, our approach emphasizes the interaction between THz photons and plant cellular structures. Several key findings emerged: 1) Clubroot infection exerts systemic effects beyond the roots, significantly altering the physiological and structural properties of leaf tissues; 2) Clubroot induces a frequency-dependent shift in refractive index – lower at low frequencies and higher at high frequencies – indicating both impaired water transport and altered metabolic states in infected tissues; 3) THz interaction with root tissues is stronger and more complex than with leaves, as evidenced by pronounced changes in amplitude, refractive index, and absorption maps; 4) The characteristic "swelling" of infected roots reflects internal tissue disruption rather than increased water content, consistent with a decrease in refractive index across the entire THz range. Moreover, absorption spectra of infected roots exhibit disordered patterns and distinct peaks at low frequencies, contrasting with smoother spectra observed in leaves. Finally, we proposed a novel physics-constrained neural network to extract the feature in THz-TDS. A comprehensive evaluation, including time-domain signals, amplitude and phase images, refractive index and absorption coefficient maps, and principal component analysis, are employed on healthy and diseased plants. Together, these results demonstrate that THz-TDS is a powerful technique for detecting plant diseases by capturing subtle changes in tissue structure, water distribution, and biochemical composition. Beyond clubroot, this method holds promise for diagnosing other biotic and abiotic stressors such as drought, nutrient deficiencies, and fungal infections. As a non-destructive and preparation-free approach, THz-TDS provides a valuable tool for early disease detection and monitoring, with potential applications in precision agriculture and plant health management.

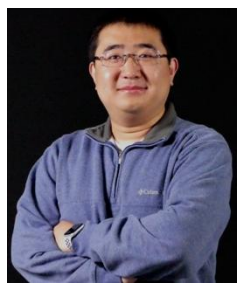
**Jiaxu Wu** is a PhD candidate in the plant biology program at Université Laval, Canada. Currently, his research focuses on understanding the clubroot resistance in canola through a multidisciplinary approach that encompasses genetics, biochemistry, and bioinformatics. The primary objective of my project is to assist the Canadian canola industry in effectively managing clubroot disease while maintaining sustainability. He holds an M.Sc. degree from Memorial University of Newfoundland, Canada, which he earned in 2022. At Memorial University, he investigated the mechanisms of plasma membrane lipidome remodelling in maize cold tolerance.

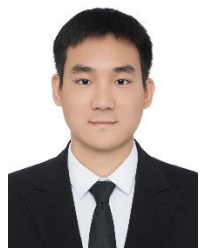
**Yubin Zhang** received the B.Eng. degree in safety engineering in 2021 from the China University of Petroleum, Qingdao, China, where he is currently working toward the Ph.D. degree in safety science and engineering. From February 2025 to February 2026, he is a research intern with the Department of Electrical and Computer Engineering, Université Laval, Québec, Canada, under the supervision of Prof. Xavier Maldague.

His research interests include non-destructive testing, infrared thermography, ultrasound, acoustic emission, deep learning, terahertz time-domain spectroscopy, and industrial data modeling.

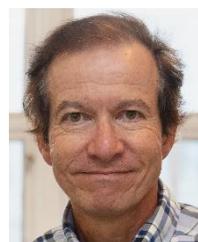
**Xavier Maldague** P.Eng., Ph.D. is full professor at the Department of Electrical and Computing Engineering, Université Laval, Québec City, Canada. He has trained over 50 graduate students (M.Sc. and Ph.D.) and contributed to over 400 publications. His research interests are in infrared thermography, NonDestructive Evaluation (NDE) techniques and vision / digital systems for industrial inspection. He is an honorary fellow of the Indian Society of Nondestructive Testing, fellow of the Canadian Engineering Institute, Canadian Institute for NonDestructive Evaluation, American Society of NonDestructive Testing. In 2019 he was bestowed a Doctor Honoris Causa in Infrared Thermography from University of Antwerp (Belguim).

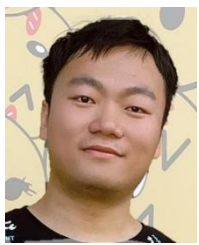
**Pengfei Zhu** received the B.Eng. degree in engineering mechanics from North University of China, Taiyuan, China, in 2019, and the M.Eng. degree in solid mechanics from Ningbo University, Ningbo, China, in 2022, and the Ph.D. degree in electrical engineering from Université Laval, Québec, Canada, in 2025.

He is an Adolf Martens Fellow in Bundesanstalt für Materialforschung und – prüfung (BAM). His research interests include non-destructive testing, infrared thermography, deep learning, terahertz time-domain spectroscopy, and photothermal coherence tomography.